\documentclass[conference]{IEEEtran}
\usepackage[pdftex]{graphicx}

\ifCLASSINFOpdf
\else
\fi
\begin{document}
%
\title{A new high speed, Ultrascale+ based board for the ATLAS jet calorimeter trigger system }

\author{\IEEEauthorblockN{B.~Bauss, V.~B{\"u}cher, J.~Damp, R.~Degele, H.~Herr, C.~Kahra, S.~Rave, E.~Rocco, U.~Sch{\"a}fer, J.~Souza, D.B.~Ta,\\S.~Tapprogge, M.~Weirich}
\IEEEauthorblockA{Institut f{\"u}r Physik, Johannes Gutenberg Universit{\"a}t, Mainz Germany}
\and
\IEEEauthorblockN{A.~Brogna}
\IEEEauthorblockA{Detector Lab, PRISMA cluster of Excellence, Mainz Germany}}


%


\maketitle

\IEEEpeerreviewmaketitle

\section{Introduction}
After Long Shutdown 2 (LS2) of the accelerator complex at CERN, the Large Hadron Collider (LHC) will restart for Run3 with an increased luminosity, $\sim$2.5 $\times$ 10$^{34}$ cm$^{-2}$s$^{-1}$. The ATLAS experiment \cite{atlas} has planned a detector upgrade for the Phase-I period (from 2019 to 2023) to address the new challenging accelerator machine conditions and to maintain the sensitivity to electroweak physics without being affected by the increased number of pile-up events.
This contribution describes the jFEX part of the upgrade of the ATLAS Level-1 Calorimeter Trigger System \cite{ATLAS_tdr}. This upgrade consists of three new Feature Extractor (FEX) systems (electron FEX (eFEX) \cite{eFEX}, jet FEX (jFEX) and global FEX (gFEX) \cite{gFEX}) which differ in the physics objects used for the trigger selection and process digitised data from the calorimeters distributed by a new optical plant (FOX). At the start of LHC Run 3, during the commissioning of these new systems, the current legacy system will initially run in parallel before later being retired \footnote{From ATL-DAQ-PROC-2018-004. Published with permission by CERN.}.

\section{jFEX requirements and features}

The jFEX board will identify jet and tau candidates and calculate the transverse energy sum, $\Sigma E_{T}$, and the missing transverse energy, $E_{T}^{miss}$. It will receive data from the central and forward calorimeter with various granularities over the $\eta$ range of $\pm$ 4.9. The input bandwidth has been maximised so that the jFEX can exploit the calorimeter information at a granularity of 0.1 $\times$ 0.1 in $\eta \times \phi$ whilst maintaining the capability of large area jet clustering. Several algorithms for the jet identification will run on the jFEX, with the current baseline algorithm being the "\textit{Sliding window}", which identifies the local maximum and for each local maximum sums the surrounding cells in a round window of R=0.4.


\begin{figure}[!t]
\centering
\includegraphics[width=3.5in]{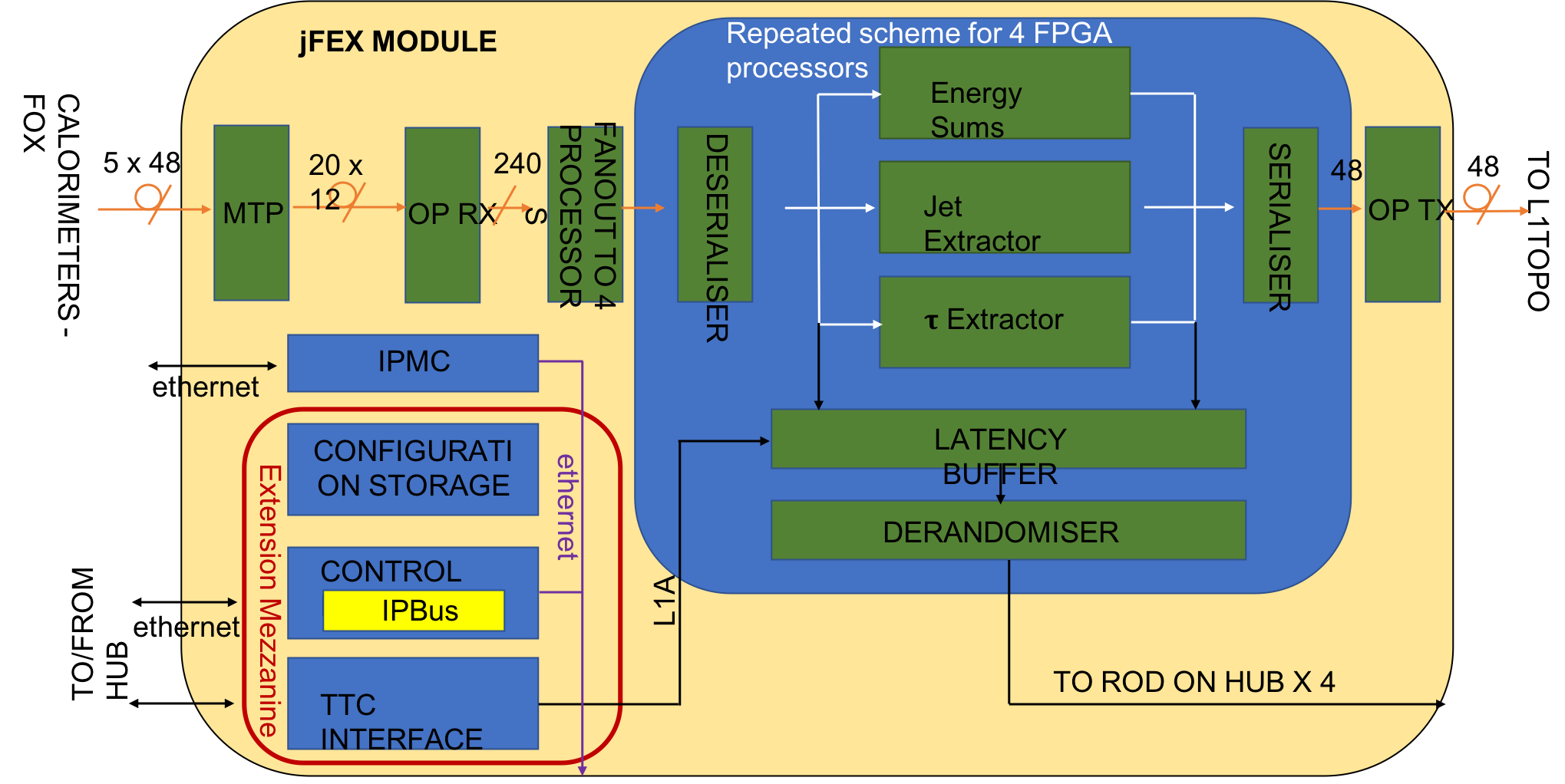}
\caption{jFEX block diagram. The sketch shows the functionality of one of the four FPGAs and on bottom left the functionalities of the extension mezzanine. Data and module monitoring are performed through the connection to the HUB and the ROD via backplane.}
\label{fig_block_diagram}
\end{figure}

The jFEX functionalities are shown in a block diagram in Fig.~\ref{fig_block_diagram}. The digitised calorimeter data are fed into the jFEX through the Fibre Optical Plant (FOX) and then received on the board through the opto-electrical devices, miniPODs \cite{avago}.
Each module has four FPGAs: each FPGA covers an area of 2.4 $\times$ 3.2 in $\eta \times \phi$, with the core area of 0.8 $\times$ 1.6. The module covers the whole ring in $\phi$ with overlapping regions between processors in $\eta$ and $\phi$ duplicated on board using PMA loopback. The results of jet identification and tau algorithms and the calculation of transverse energy sum and missing transverse energy are sent to the Level-1 Topological Processor board \cite{topo} as Trigger Objects (TOBs).

The board control is located on a mezzanine to allow flexibility in upgrading control functions and components without affecting the main board. This design choice guarantees smooth and reliable board operation and compatibility with the surrounding trigger system for the next twenty years of data taking.

The data readout checks and external communications are performed via the ROD \cite{ATLAS_tdr} and the HUB \cite{ATLAS_tdr}.


\subsection{jFEX design challenges}

The FPGA choice was driven by the high input bandwidth and large processing power required for the complex algorithms that will be implemented. The Xilinx Ultrascale+ XCUV9P-2FLGA2577 device \cite{xilinx} meets the requirement of large number of Multi Gigabit Transceivers (MGT) and of the processing power. This device has 120 MGTs, capable to handle up to 3.6 Tb/s input bandwidth. The jFEX board hosts four Xilinx Ultrascale+ and twenty-four miniPODs, twenty receivers and four transmitters. The jFEX board is an ATCA-based \cite{atca} board therefore the routing space available is quite limited taking into account all the components on it.

The challenges faced during the design of the board are summarised below:

\begin{itemize}

   \item signal integrity

   \item FPGA power consumption

   \item FPGA power dissipation

\end{itemize}

\section{jFEX Prototype}

\begin{figure}[!h]
\centering
\includegraphics[width=3.0in]{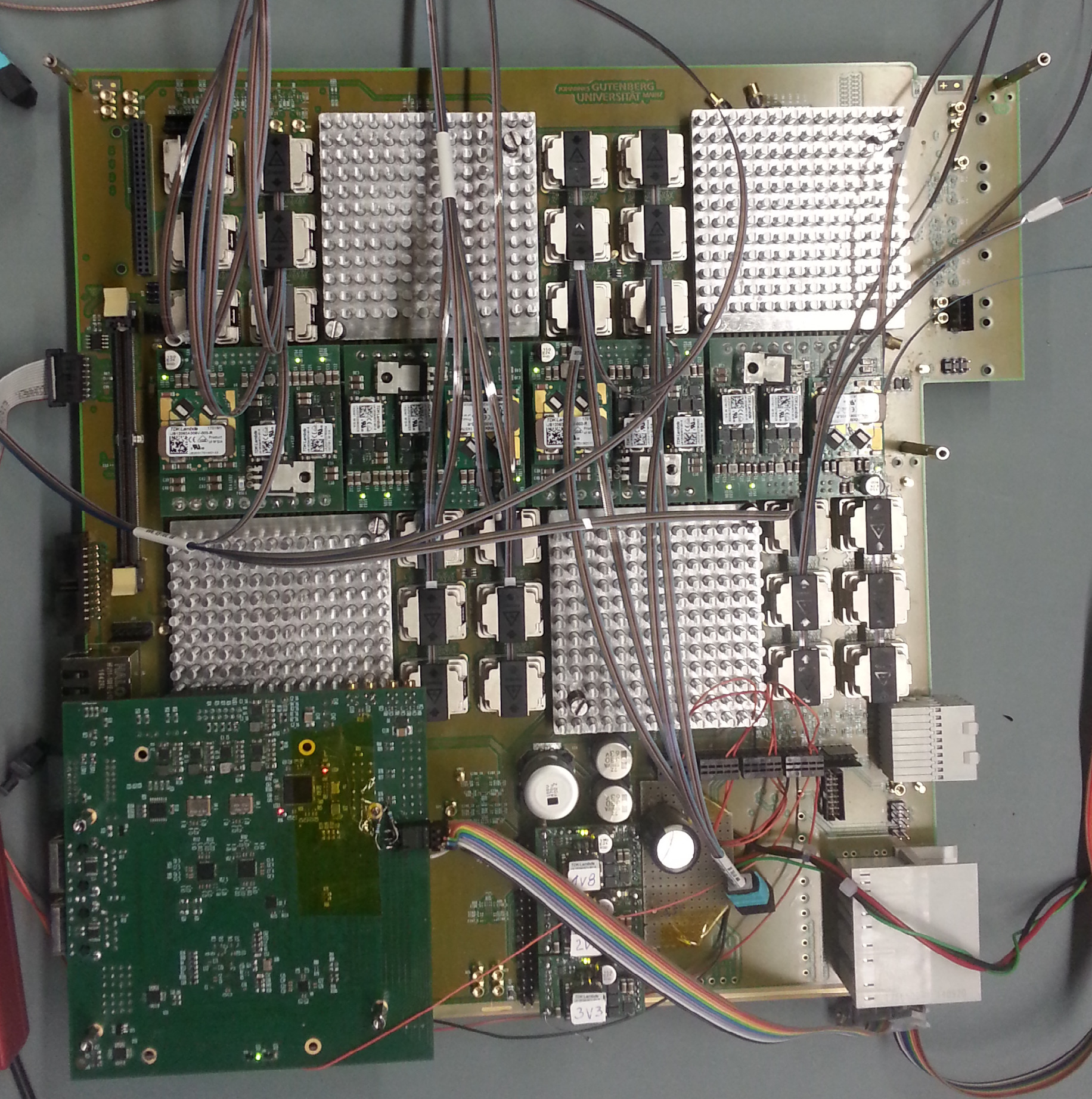}
\caption{Picture of the final jFEX prototype.}
\label{fig_prototype}
\end{figure}

The first jFEX prototype was delivered at the end of 2016 and it was equipped with one Xilinx Ultrascale, XCUV190-2FLGA2577 (pin compatible with the Xilinx Ultrascale+ XCUV9P-2FLGA2577) and the six closest opto-electrical devices.
The final jFEX prototype, see Fig.~\ref{fig_prototype} was delivered in November 2017. It is fully equipped with four Xilinx Ultrascale+ XCUV9P-2FLGA2577. 

The prototype characterisation was performed in two different test campaigns: at Mainz University and in the integrated test at CERN. 

\section{Test Results}

Once the final prototype board was received, basic tests were immediately performed at the Mainz University Laboratory. A first validation of all the optical and electrical links was done. In February 2018, a full board validation has been performed at the integrated test at CERN where the measurement conditions are very close to the final ones.

To exercise all the jFEX inputs two FEX Test Modules (FTM) \cite{ATLAS_tdr} were used and the readout links were also verified using the HUB+ROD modules.

\begin{figure}[!t]
\centering
\includegraphics[width=3.7in]{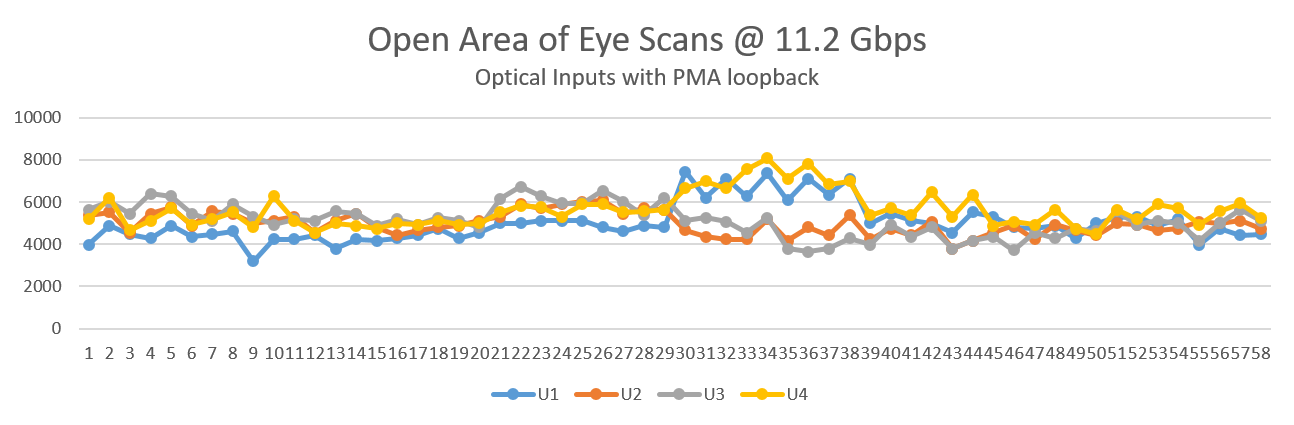}
\caption{Open Area of Eye Diagram of all the jFEX input links duplicated via PMA loopback for the four processors. On the x axis the number of the channels  per processor are reported while on the y axes the area of the eye diagram in a.u..}
\label{all_links}
\end{figure}

Fig.~\ref{all_links} shows the open areas of the eye diagrams for the duplicated data links taken by Xilinx IBERT IPCore. PRBS31 data was sent on optical fibres from FTM (48 links) and from a jFEX transmitter (10 links) to a first jFEX processor, where it was duplicated inside the MGTs (far-end PMA loopback), and sent electrically to a second processor, where the eye diagrams were measured. In this measurement all input links were running concurrently and a Bit Error Rate (BER) of 10$^{-15}$ was reached.

A dedicated test to assess the quality of the clock on board was performed. A spectrum analyser was used for the clock jitter measurement as the Ultrascale+ clock jitter is specified in the frequency domain.

The proper thermal behaviour of the regulators and of the FPGAs was checked by monitoring them in several conditions of FPGA load and with thermal camera measurements. 


\section{Firmware}

\begin{figure}[!t]
\centering
\includegraphics[width=3.5in]{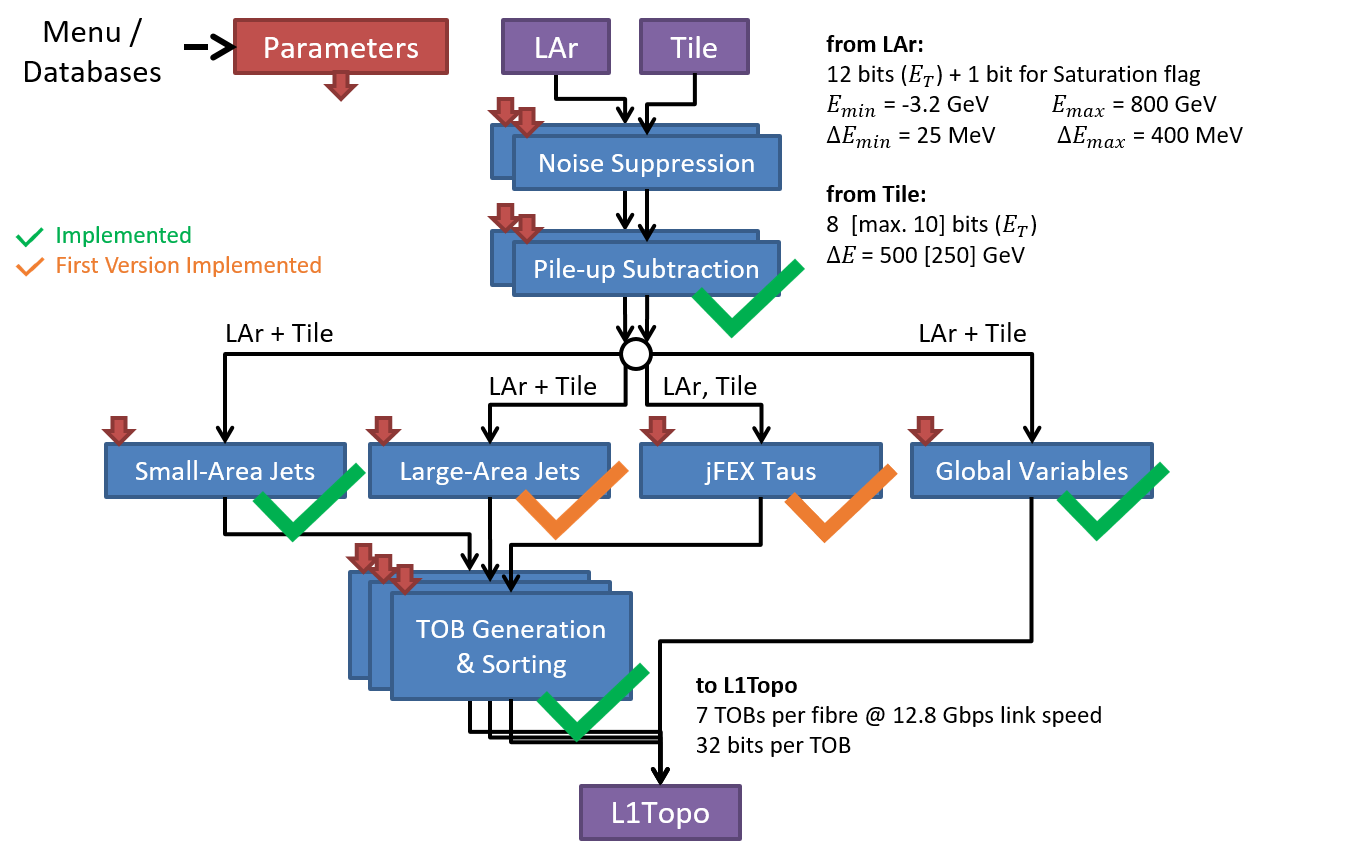}
\caption{jFEX Algorithm Block Diagram.}
\label{jfex_fw}
\end{figure}

The jFEX will identify jets, taus and global variable with a full calorimeter coverage ($\mid \eta \mid <$ 4.9). The baseline algorithms shown in Fig.~\ref{jfex_fw} are already implemented (a preliminary one in the forward region)
and the FPGA resource usage are 18\% LUT in the central region and 47\% in the forward region. The forward region, namely for 3.1 $< \mid \eta \mid <$4.9, has an irregular mapping in ($\eta$, $\phi$) therefore more FPGA resources are needed.

The output of the algorithms are sorted internally on the FPGA before being sent to L1Topo.


\section{Conclusions}

The jFEX system is part of the Phase-I upgrade to the ATLAS Level-1 Calorimeter Trigger.  The jFEX final prototype has been extensively tested and validated during two test campaigns at Mainz and CERN.
The jFEX processes the data in real-time and without any buffering within a latency budget of $<$390 ns with fixed MGTs latency including the data duplication on board via PMA loopback.  

The jFEX pre-production is imminent and the full production will be completed at the end of the current year. The jFEX system will be installed and commissioned within the ATLAS detector during LS2 before the restart of the LHC for Run 3.







%

\end{document}